\documentclass[trackchanges]{aastex701}
\usepackage{amsmath}
\usepackage{natbib}

\journalinfo{}

\begin{document}

\title{Specific Star Formation Rate Enhancement Across the Galaxy Merger Sequence: Insights from Citizen Science Classifications}

\author[0009-0008-8109-6871]{Jacob Lee}
\affiliation{Minnetonka High School, 18301 Highway 7, Minnetonka, MN 55345}
\affiliation{Minnesota Institute for Astrophysics, University of Minnesota, 116 Church Street SE, Minneapolis, MN 55455, USA}
\email[show]{lee05337@umn.edu}

\author[0000-0003-1767-6421]{Alexandra Le Reste}
\affiliation{Minnesota Institute for Astrophysics, University of Minnesota, 116 Church Street SE, Minneapolis, MN 55455, USA}
\email[show]{alereste@umn.edu}

\author[0000-0002-9136-8876]{Claudia Scarlata}
\affiliation{Minnesota Institute for Astrophysics, University of Minnesota, 116 Church Street SE, Minneapolis, MN 55455, USA}
\email[]{mscarlat@umn.edu}

\author[0000-0002-6016-300X]{Kameswara Bharadwaj Mantha}
\affiliation{Department of Physics \& Astronomy, University of Missouri Kansas City, 5110 Rockhill Road, Kansas City, MO, 64110.}
\affiliation{Missouri Institute for Defense \& Energy (MIDE), 5110 Rockhill Road, Kansas City, MO, 64110}
\email[]{km4n6@umkc.edu}

\begin{abstract}
We present an analysis of specific star formation rates (sSFR) across the galaxy merger sequence using visual classifications from the Zooniverse citizen science project ``Cosmic Disco: Characterizing Galaxy Collisions''. Our sample comprises 4,884 galaxy systems pre-selected as merger candidates from SDSS DR17 ($0.01 < z < 0.05$, $M_* > 10^{8.5}M_\odot$) using Zoobot, of which 3690 were classified as mergers spanning pre-interaction through post-coalescence stages by citizen scientist volunteers. We find a weak but statistically significant positive correlation between $\log(\mathrm{sSFR})$ and visual merger stage ($r = 0.161$, $p = 7.23 \times 10^{-23}$), with a best-fit relation $\log\left(sSFR\right)=(0.148\pm0.015)\, S_{\rm Merg}-(1.865\pm0.038)$. The large RMS scatter (0.661 dex) reflects visual merger stages capturing wide merger timescales, and our results corroborate previous findings of increasing SFR enhancement with merger progression. This work shows that citizen science is a viable complement to automated and pair-based approaches to evaluate timescales for galaxies across the merger sequence.
\end{abstract}


\section{Introduction}

Galaxy mergers are thought to be a key driver of galaxy evolution, influencing morphological transformation, star formation rates (SFR), formation of starburst galaxies, and active galactic nucleus (AGN) activity through cosmic time \citep{Conselice2021}. However, it is hard to observationally identify mergers, which is further complicated by their long evolutionary timescales that can span over a billion years. Robustly assessing the impact of mergers on specific galaxy properties requires large samples of galaxies at known merger stages.

Two main methods are used for identifying galaxy mergers: we can look for morphologically distorted galaxies, or identify galaxies in pairs, which have high chances of merging in the future \citep{Conselice2021}. While pair identification is often used, it is limited to galaxies with known redshifts, and pairs cannot characterize the critical post-merger regime where many physical processes are predicted to peak \citep{Ellison2013}. Visual identification is also widely used for merger characterization, leveraging morphological classifiers and citizen science classifications \citep{Darg2010}. For decades now, astronomers have been characterizing the timescales of merger interactions visually \citep{LeFevre2000}. However, this has generally been done on a relatively small number of galaxies by few people, with limited data availability. Here, we explore visual classifications from citizen science project \textit{Cosmic Disco: Characterizing Galaxy Collisions}\footnote{\url{https://www.zooniverse.org/projects/kbmantha/cosmic-disco-characterizing-galaxy-collisions}}; hereafter referred to as Cosmic Disco. Cosmic Disco aims to identify the timescale of mergers by aggregating classifications and votes from thousands of citizen scientists, providing a comprehensive library of galaxy mergers with available merger timescales (A. Le Reste \& K. B. Mantha et al., in prep). This study uses the merger stages identified in Cosmic Disco to analyze how the specific star formation rate (sSFR) evolves as a function of merger stage across the merger sequence, including mergers closer than 5 kpc and post-coalescence.

\section{Methodology}

Cosmic Disco galaxies were selected from the Sloan Digital Sky Survey (SDSS) Data Release 17 \citep{Abdurrouf2022} to have available spectral line fluxes from the MPA-JHU catalog \citep{Brinchmann2004}. These were cross-matched with the DESI Legacy Imaging Surveys \citep{Dey2019}, targeting systems with redshift $0.01<z<0.05$ and stellar-mass $>10^{8.5}M_{\odot}$ (88216 galaxies). These were then filtered using the Zoobot model trained on the Galaxy Zoo project \citep{Walmsley2023}. Galaxies with Zoobot score $f_{\rm merger}>0.1$ were pre-selected as possible mergers, yielding a total of 7,244 galaxies to be classified in Cosmic Disco. Additionally, 41,472 systems with high emission line S/N and AGNs excluded were classified as non-mergers with Zoobot ($f_{merger}<0.05$) and non-pairs using projected separation and velocity separation criteria $r_p>150\,\rm{kpc}$ and $\Delta_v>1000\,\rm{km}.\rm{s}^{-1}$. These were not presented to Cosmic Disco volunteers; they constitute the SDSS control sample.

The five choices of classifications in Cosmic Disco were ``Not a merger'', ``Pre-Interaction'' (score = 1), ``Post-Interaction'' (score = 2), ``Near-Coalescence'' (score = 3), and ``Post-Coalescence'' (score = 4). We define the visual merger fraction $P_{merg}$ as the fraction of votes cast for any of the four merger categories. Galaxies with $P_{\rm merg} < 0.5$ were given the non-merger label (1766 galaxies), while systems with $P_{\rm merg}\geq0.5$ were designated as mergers (5478 galaxies). Each galaxy was classified by an average of 18 volunteers. We average the scores for mergers, designating a mean stage $S_{\rm Merg}$, characterizing a continuum of timescales from pre-interaction ($S_{\rm Merg}$=1) to post-coalescence ($S_{\rm Merg}$=4).

For this study, galaxies with low emission line flux S/N were removed (S/N$<$3 in H$\alpha$ and H$\beta$, 768 galaxies). AGNs were removed using the \citet{Kauffmann2003} criterion (1264 galaxies). Some galaxies had multiple spectra: only the spectrum with highest emission line S/N was kept. We corrected line fluxes for internal dust attenuation using the observed Balmer decrement assuming Case B recombination ratio $(H\alpha/H\beta = 2.86)$ and \citet{Calzetti2000} attenuation law with $R_V = 4.05.$ Dust-corrected $H\alpha$ luminosities were converted to SFR using the \citet{Murphy2011} calibration, assuming a \citet{Kroupa2001} IMF $(SFR = 5.37 \times 10^{-42} L_{H\alpha}).$ Specific star formation rates were calculated by normalizing SFR to stellar mass. Finally, quality cuts excluded galaxies with anomalously high dust-corrected $H\alpha$ fluxes ($\geq10^{-12}\ \mathrm{erg\ s^{-1}\ cm^{-2}}$), due to anomalous SDSS r-band photometric correction factor ”spectofiber” (2 mergers, 1 non-merger, 16 SDSS control). The final sample consists in 3,690 mergers, 1,194 Cosmic Disco non-merger, and 41,456 SDSS control galaxies.

\begin{figure*}[t!]
\plotone{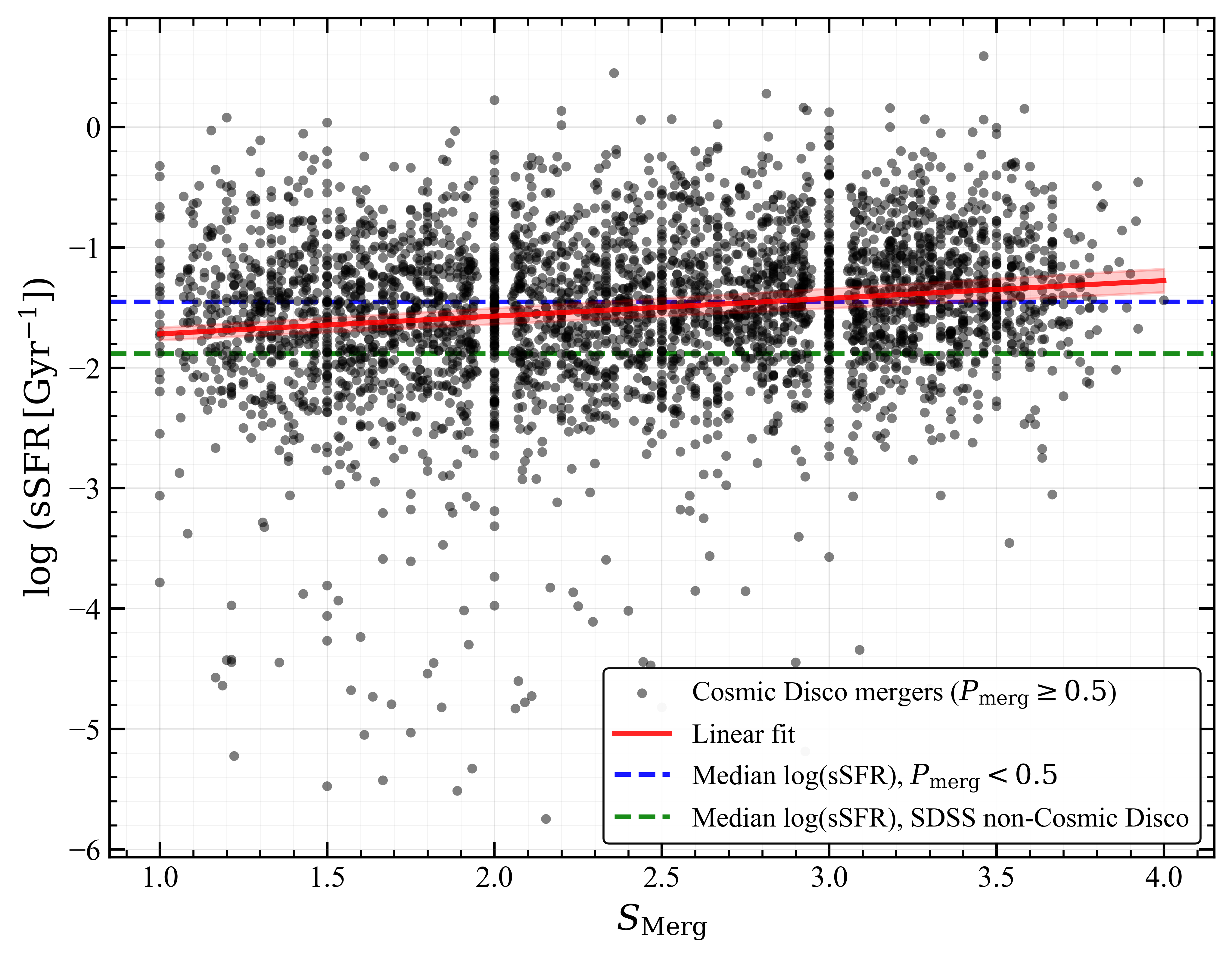}
\caption{Log(sSFR) versus $S_{\rm Merg}$ for Cosmic Disco galaxies with $P_{\rm merg} \geq 0.5$. Merger stages trace galaxies from pre-interaction ($S_{\rm Merg}=1$) to post-coalescence ($S_{\rm Merg}=4$). We find a weak but statistically significant correlation, with linear fit shown in red and $1\sigma$ uncertainty shown as the shaded region. Median $\log(\mathrm{sSFR})$ for Cosmic Disco non-mergers ($P_{\rm merg} < 0.5$; blue dashed) and SDSS control galaxies (green dashed) are shown for comparison.}
\end{figure*}

\section{Results}

We show log(sSFR) as a function of $S_{\rm Merg}$ in Figure 1. We find a weak but statistically significant correlation fit by: $$\log\left(sSFR\right)=(0.148 \pm 0.015)\, S_{\rm Merg}-(1.865 \pm 0.038)$$ The distribution of $\log(\mathrm{sSFR})$ is broad across all merger stages, with an RMS scatter of $0.661$ dex around the best-fit relation, indicating that galaxies span a wide range of star formation activity at any given merger stage. The Pearson correlation coefficient ($r = 0.161$) between $\log(\mathrm{sSFR})$ and merger stage $S_{\rm Merg}$ is modest but statistically significant ($p = 7.23 \times 10^{-23}$), suggesting that while individual galaxy-to-galaxy variation is large, there is a general trend of increasing sSFR as galaxies progress through the merger sequence from pre-interaction to post-coalescence stages. The median $\log(\mathrm{sSFR})$ of Cosmic Disco non-mergers (0.035 Gyr$\,^{-1}$) is elevated relative to the SDSS control sample (0.013 Gyr$\,^{-1}$), suggesting that some galaxies pre-selected by Zoobot but visually classified as non-mergers may still be interacting. Results are consistent with pair separation studies that find that SFR enhancement increases with decreasing projected separation \citep{Patton2013}. The large scatter and modest correlation are consistent with findings that statistically the SFR only increases by a limited factor for individual merging galaxies \citep{Pan2019}, suggesting that strong starbursts are not a universal outcome across merger timescales.

\section{Conclusions}

Using citizen science Cosmic Disco classifications, we find sSFR rises across the merger sequence with large scatter, consistent with mergers statistically enhancing rather than uniformly triggering star formation. A forthcoming paper (Le Reste \& Mantha et al., in prep) will extend this framework to additional galaxy properties.

\bibliography{References}{}
\bibliographystyle{aasjournalv7}
\end{document}